\documentclass[conference]{IEEEtran}
\IEEEoverridecommandlockouts
\usepackage{threeparttable}
\usepackage{algorithm}
\usepackage{algorithmicx}  
\usepackage{algpseudocode}
\usepackage{float}
\usepackage{cite}
\usepackage{amsmath,amssymb,amsfonts}
\usepackage{inputenc}
\usepackage{graphicx}
\usepackage{subfigure}
\usepackage{textcomp}
\usepackage{xcolor}
\usepackage{url}
\usepackage{multirow}
\usepackage{booktabs}
\usepackage{xcolor}
\def\BibTeX{{\rm B\kern-.05em{\sc i\kern-.025em b}\kern-.08em
    T\kern-.1667em\lower.7ex\hbox{E}\kern-.125emX}}

\begin{document}

\title{Malware Detection with LSTM using Opcode Language}
\author{
    \IEEEauthorblockN{
    Renjie Lu, 
    }
    \IEEEauthorblockA{University of Chinese Academy of Sciences}
    \IEEEauthorblockA{Beijing, China}
    \IEEEauthorblockA{Email: {lurenjie17@mails.ucas.ac.cn}
     \thanks{Corresponding author: lurenjie17@mails.ucas.ac.cn}
    }
 }

\maketitle

\begin{abstract}
Nowadays, with the booming development of Internet and software industry, more and more malware variants are designed to perform various malicious activities. Traditional signature-based detection methods can not detect variants of malware. In addition, most behavior-based methods require a secure and isolated environment to perform malware detection, which is vulnerable to be contaminated. In this paper, similar to natural language processing, we propose a novel and efficient approach to perform static malware analysis, which can automatically learn the opcode sequence patterns of malware. We propose modeling malware as a language and assess the feasibility of this approach. First, We use the disassembly tool IDA Pro to obtain opcode sequence of malware. Then the word embedding technique is used to learn the feature vector representation of opcode. Finally, we propose a two-stage LSTM model for malware detection, which use two LSTM layers and one mean-pooling layer to obtain the feature representations of opcode sequences of malwares. We perform experiments on the dataset that includes 969 malware and 123 benign files. In terms of malware detection and malware classification, the evaluation results show our proposed method can achieve average AUC of 0.99 and average AUC of 0.987 in best case, respectively.
\end{abstract}


\section{Introduction}
Malicious software is referred to as malware, which is designed to perform various malicious activities, such as stealing private information, gaining root authority, disabling targeted host and so on. Meanwhile, with the booming development of Internet and software industry, more and more variants of malware are emerging and almost everywhere. According to a 2018 McAfee threats report \cite{r1}, the total number of malware samples has grown almost 34\% over the past quarters to more than 774 million samples. It can be seen that the number of malware continues to increase. Hence, malware detection is always a attractive and meaningful issue.

A large number of research have been published on how to detect malware. Malware detection can be simply considered as a binary classification problem, and traditional anti-virus software usually relies on static signature-based detection method \cite{r2}, which has a significant limitation. some minor changes in malware can change the signature, so more malware could easily evade signature-based detection by encrypting, obfuscating or packing. Meanwhile, the zero-day malware can also evade this detection approach. The dynamic analysis is not susceptible to code obfuscation techniques \cite{r3}, so it is a more effective malware detection method. Dynamic behavior-based malware detection methods \cite{r4}\cite{r5} usually need a secure and controlled environment, such as virtual machine, simulator, sandbox, etc. Then the behavior analysis is performed by using the interaction information with the environment such as API calls and DLL calls. Although these techniques have been widely studied, they have also been confirmed to be less efficient enough when applied to large dataset \cite{r6}. Dynamic behavior-based malware detection methods are quite time-consuming and require considerable attention to protect the operating environment from contaminated. 

At present, a number of malware detection methods combined with machine learning techniques have been proposed. Reference \cite{r7} first proposed a malware detection method using data mining technique, which use three different types of static features: PE header, string sequence, and byte sequence. Kolter and Maloof \cite{r8} proposed to use n-gram instead of byte sequence and compared the performance of naive bayes, decision trees, support vector machines for malware detection. Later, artificial neural network \cite{r9}\cite{r10} were also used for malware detection. Meanwhile, there are also some novel ideas for malware detection. Both \cite{r11} and \cite{r12} utilize the technique of image processing to detect malware. In terms of malware detection, the previous works have achieved good enough performance. However, most of these methods manually extract malware features which are used to train a machine learning classifier.

To reduce the cost of artificial feature engineering, in this paper, we propose a novel and efficient method to detect whether a Windows executable file is malware. First, we use the disassembly tool IDA Pro to obtain the assembly format file of all executable files. Next, we develop an algorithm to extract opcode sequence from each assembly format file. Then, similar to natural language processing (NLP), word embedding technology \cite{r13} is used to learn the feature vector representation of opcode, and long-short term memory (LSTM) \cite{r14} is used to automatically learn opcode sequence patterns of malware. To increase invariance of the local feature representation, we also introduce a mean-pooling layer after second LSTM layer. To verify the effectiveness of our proposed method, we make a series of experiments on the dataset that includes 969 malwares and 123 benign files. In the experimental section, we evaluate the effect of the second LSTM layer on malware detection performance, and we also make detailed performance comparison with other related work.
In terms of malware detection and malware classification, the evaluation result shows our proposed method can achieve average AUC of 0.99 and average AUC of 0.987, respectively.

In summary, we make the following contributions in this paper:
\begin{itemize}
\item We present a novel and efficient malware detection approach, which makes use of word embedding technology and LSTM to automatically learn the opcode sequence patterns of malwares. It can greatly reduce the cost of artificial feature engineering.
\item We propose and implement a two-stage LSTM model for malware detection, which use two LSTM layers and one mean-pooling layer to automatically obtain the comprehensive feature representation of malware.
\item We make a series of evaluation experiment including malware detection and malware classification. The exprimental results demonstrate the effectiveness of our proposed method.
\end{itemize}

The rest of this paper is organized as follows. Related work on neural network is discussed in Section II. Section III describes the proposed malware detection framework. Experiment and evaluation are presented in Section IV. Section V concludes the paper and discusses the future work.

\section{Related Work}

\subsection{Word Embedding}
Recurrent Neural Network Based Language Model (RNNLM) \cite{r15} is the language model using recurrent neural network, which can predict the next word from previous input. Later, Mikolov \cite{r13} proposed the CBOW and Skip-gram language model in efficient estimation of word representations in vector space. CBOW model can predict current word by the given context. In contrast, Skip-gram model can predict context by the given current word. Then, each word is converted to feature vector which store the semantic information of word, and the correlation between words can also be calculated using this feature vector.

\subsection{Recurrent Neural Network}
Neural network (NN) is a kind of mathematical model which is consisted of many neuron layers. Recurrent Neural Network (RNN) is a typical structure of NN, and it has a special memory unit which can retain the state information of previous hidden layer. 
RNN shows good results in various fields which use sequential data such as natural language processing and speech recognition. A large amount of research works using RNN for malware detection has been published. EI-Bakry \cite{r16} proposed that a time delay neural networks could be used for malware classification, but the paper did not carry out any experiments to validate the idea. Pascanu et al. \cite{r17} proposed a malware detection method using RNN. However, \cite{r17} uses API calls as the original feature for malware detection. Tobiyama et al. \cite{r18} proposed a malware detection method with deep neural network using process behavior. However, because of the discovery of the vanishing and exploding gradient problem, RNN became unpopular until the LSTM  was proposed. LSTM \cite{r14} is a special type of RNN, which can greatly mitigate the vanishing and exploding gradient problem. 

\section{Malware Detection Methodology}
In this section, we introduce the proposed malware detection approach in detail, which is similar to natural language processing. As shown in Figure 1, the malware detection methodology can be simply divided into the data processing stage and modeling stage. In the data processing stage, we first use the disassembly tool IDA Pro, which can resolve executable file into Intel X86 assembly format file. Next, we develop a algorithm to extract opcode sequence from each assembly format file. In the modeling stage, word embedding technology is used to learn the correlation between opcodes and to obtain the feature vector representation of each opcode. Then, a two-stage LSTM model is used to learn the opcode sequence patterns of each sample and to generate the predictive model. Finally, we use the predictive model to perform malware detection on the testing set in order to evaluate its performance.
\begin{figure*}[htbp]
\centering
\includegraphics[width=15cm,height=2.5cm]{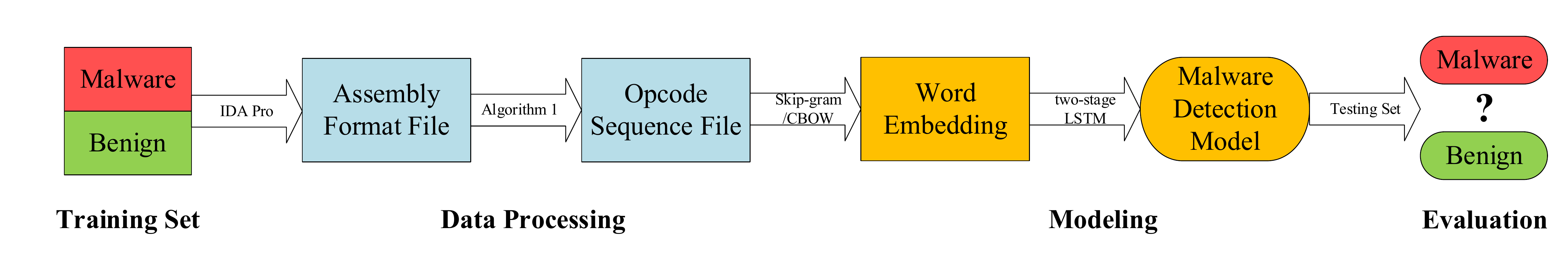}
\caption{The overview of malware detection methodology. Data processing stage includes the conversion of executable file to \emph{.asm} file and the extraction of opcode sequence. Modeling stage is consists of the word embedding and the generation of malware detection model.}
\label{fig}
\end{figure*}
\subsection{Data Processing}
In order to obtain the features of opcode sequence, we need to extract opcode sequence from each assembly format file. Typically, this type of assembly format file contains four basic predefined segments, .text segment, .idata segment, .rdata segment, and .data segment. Since only the .text segment stores program instructions and the rest of segments are data segment, we only consider the contents of .text segment to extract opcode sequence. Meanwhile, we also find some meaningless opcodes such as `dd', `db', `align' and so on. To obtain opcodes that are really beneficial for malware detection, we need to filter out these meaningless opcodes. Actually, the opcode sequence can reflect program execution logic of corresponding executable file. The pseudocode of this opcode extraction algorithm is shown in Algorithm 1.

\begin{algorithm}
\caption{Extract Opcode Sequence}
    \begin{algorithmic}[1]
        \Require Each assembly format file  
        \Ensure Corresponding opcode sequence   
        \State \textit{pattern} $\gets$ \textit{predefined matching pattern for extracting opcode}
        \State $filter \gets \left\{ `align', `dd', `db', ... \right\}$
        \State \textit{file} $\gets$ \textit{open (assembly format file)}
        \For{$eachline$ $in$ $file$}
        \If {\textit{eachline starts with '.text'}}  
        \State \textit{result} $\gets$ \textit{match (pattern, eachline)}
        \If {\textit{result is not null} \textbf{and} \textit{result not in filter}}  
        \State \textit{add result to corresponding opcode sequence} 
        \EndIf
        \EndIf
        \EndFor
    \end{algorithmic}
\end{algorithm}

\subsection{Opcode Representations in Vector Space}
Common word representations in NLP are one-hot representation, bag-of-word, or n-grams. However, the largest defect of these local representations is that any two words are isolated so that it can't reflect the semantic correlation between words. We use word embedding technique to automatically learn feature vector representation of opcode. As shown in Figure 2(a), the Skip-gram model tries to predict its context from input opcode according to the word window size. In constrast, CBOW model can predict current word by the given context as shown in Figure 2(b). If the word window size is set to i, then i is the maximum distance between the current opcode and predicted opcode.

\begin{figure}[htbp]
\centerline{\includegraphics[width=8cm,height=5.5cm]{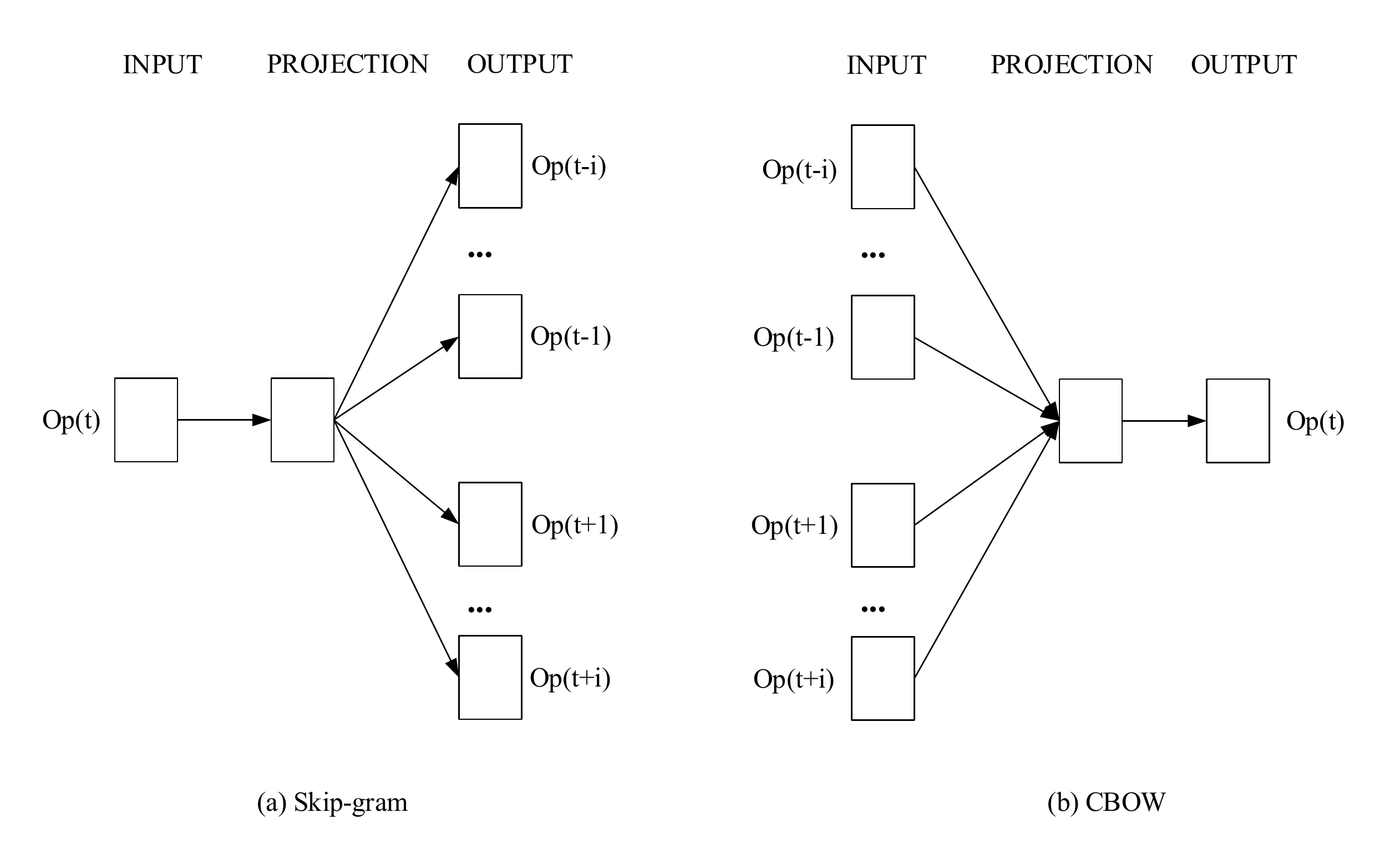}}
\caption{Two word embedding techniques: Skip-gram and CBOW.}
\label{fig}
\end{figure}

As the model can not directly process opcode in the form of string, opcode is first converted to one-hot representation. Hence, We make frequency statistics on all opcode sequence files and filter out low frequency opcodes to build an opcode vocabulary. In the end, the opcode vocabulary we created contains 391 different and valuable opcodes. Accordingly, the one-hot representation of opcode should be a 391-dimensional vector which contains only one non-zero element like [0, 0, 0, 1, 0, ..., 0], and each opcode gets a unique one-hot representation.

Then we use the Gensim Python library \cite{r19} for word embedding to obtain the feature vector representation of opcode. After multiple experimental evaluations, we set the dimension of the feature vector to 100. In this paper, we use the CBOW model to implement our proposed malware detection method. In the experimental evaluation, we will discuss the impact of different word window sizes and different word embedding techniques (Skip-gram and CBOW) on malware detection accuracy in detail. Next, we use a two-stage LSTM model to learn the comprehensive feature representation of entire opcode sequence.

\subsection{Feature Representation by LSTM}
\subsubsection{Long-short term memory }
As a typical and improved recurrent neural network, long-short term memory (LSTM) \cite{r14} is suitable for processing and predicting time series problems. LSTM model introduces a new structure called a memory cell, which is composed of three main elements: an input gate, a forget gate and an output gate, to control the transmission of information, as seen Figure 3. Because of this special structure, LSTM can alleviate the vanishing and exploding gradient problem.
\begin{figure}[htbp]
\centerline{\includegraphics[width=8cm,height=4.5cm]{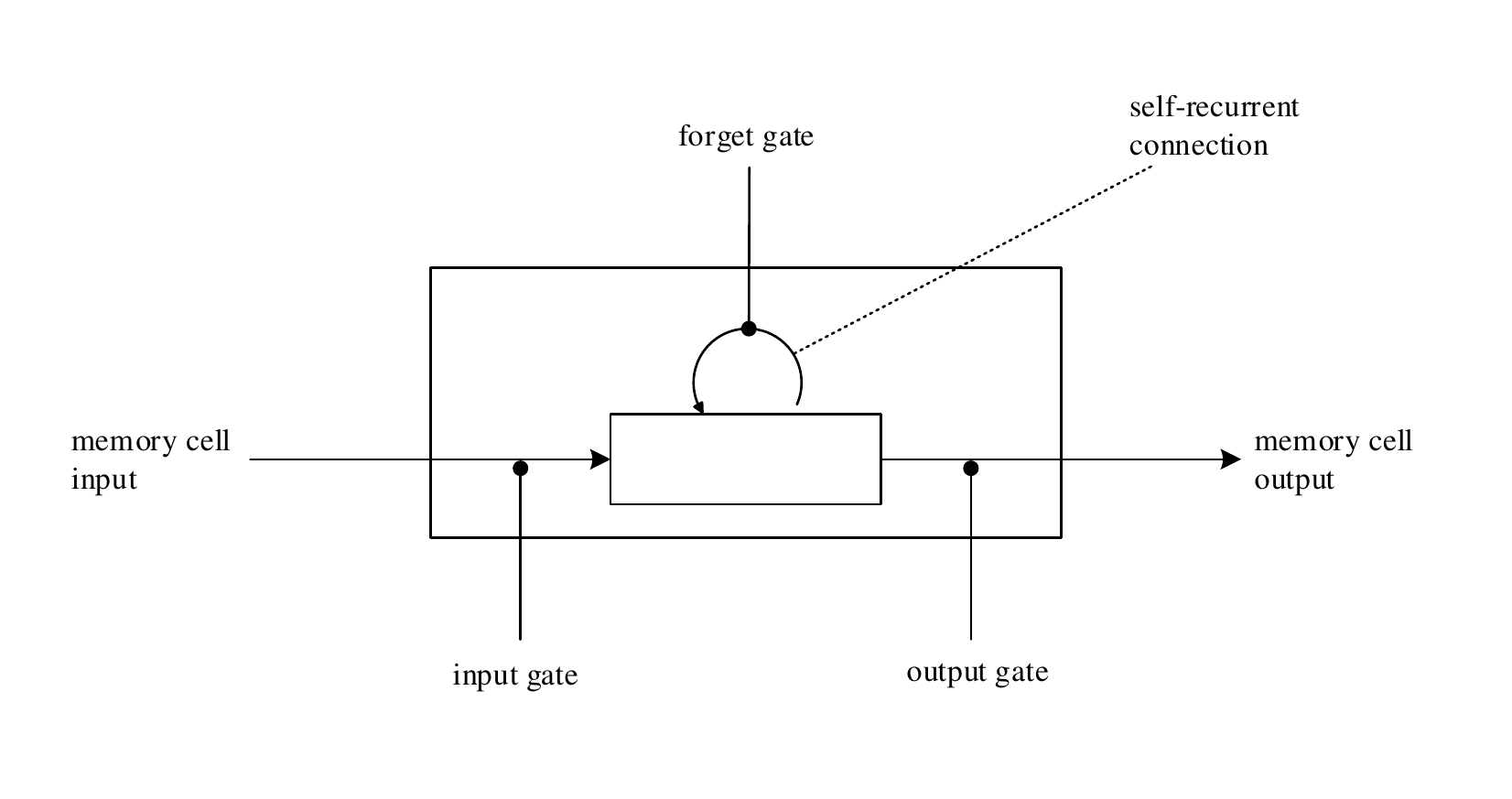}}
\caption{Illustration of a standard memory cell.}
\label{fig}
\end{figure}

Let $x_t$ denote the input to memory cell at time \emph{t}; let $W_i$, $W_f$, $W_c$, $W_o$, $U_i$, $U_f$, $U_c$, $U_o$ and $V_o$ be weight matrixes; let $b_i$, $b_f$, $b_c$ and $b_o$ be bias vectors.
Formally, the equations below describe that how a memory cell is updated at time \emph{t}.
\begin{itemize}
\item First, we calculate the value of the forget gate $f_t$, the value of the input gate $i_t$, and update the previous state of the memory cell to $\tilde{C_t}$.
\begin{gather}
f_t= \sigma(W_f x_t+U_f h_{t-1}+b_f)\\
i_t= \sigma(W_i x_t+U_i h_{t-1}+b_i)\\
\tilde{C_t}=tanh(W_c x_t+U_c h_{t-1}+b_c)
\end{gather}

\item Second, given the value of the input gate, the value of the forget gate and the value of updated state $\tilde{C_t}$, we can calculate new state of memory cell, $C_t$, at time \emph{t}.
\begin{gather}
C_t = i_t*\tilde{C_t} + f_t*C_{t-1}
\end{gather}

\item With the new state of memory cell, we can calculate the value of the output gate and the output of memory cell.
\begin{gather}
o_t=\sigma(W_o x_t+U_o h_{t-1}+V_o C_t+b_o)\\
h_t= o_t*tanh⁡(C_t)
\end{gather}
\end{itemize}

In the above formula, $\sigma$ is the logistic sigmoid function, so the value of gating vector $i_t$, $f_t$, $o_t$ are in [0,1]. \emph{tanh} is the hyperbolic tangent function and * is the pointwise multiplication operation.
 
\subsubsection{Two-stage LSTM}
We conduct detailed statistics on the frequency of each opcode that appear in the dataset. The statistical results are shown in Figure 4 and Figure 5. Figure 4 presents the average of opcodes for each type of samples and top 10 most used opcodes in the dataset. Figure 5 also presents the most frequent 10 opcode for each type of samples.
\begin{figure}[htbp]
\centerline{\includegraphics[width=6cm,height=4cm]{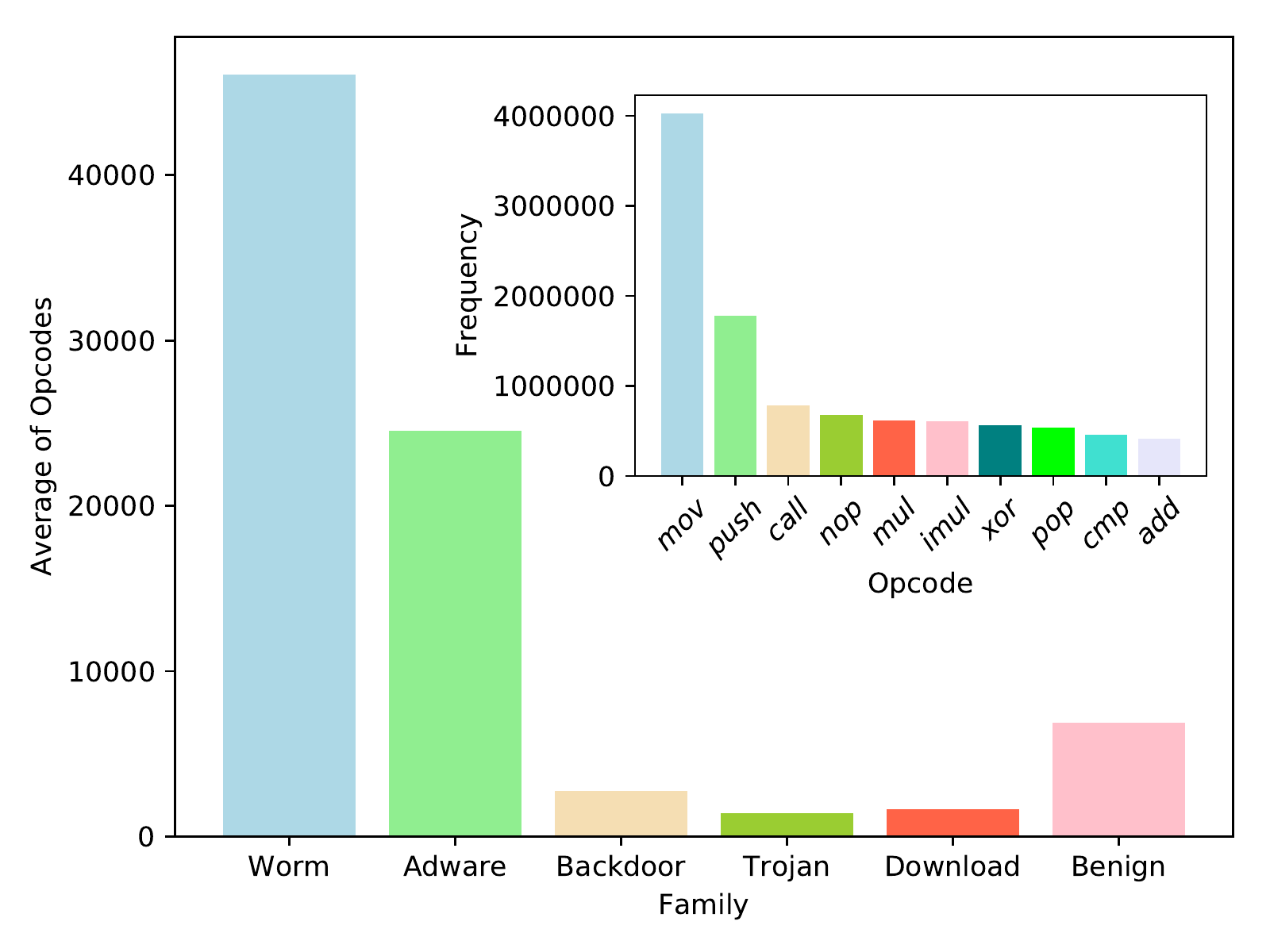}}
\caption{The average of opcodes for each type of samples and Top 10 opcodes in the dataset.}
\label{fig}
\end{figure}

\begin{figure}[htbp]
\centerline{\includegraphics[width=7cm,height=5cm]{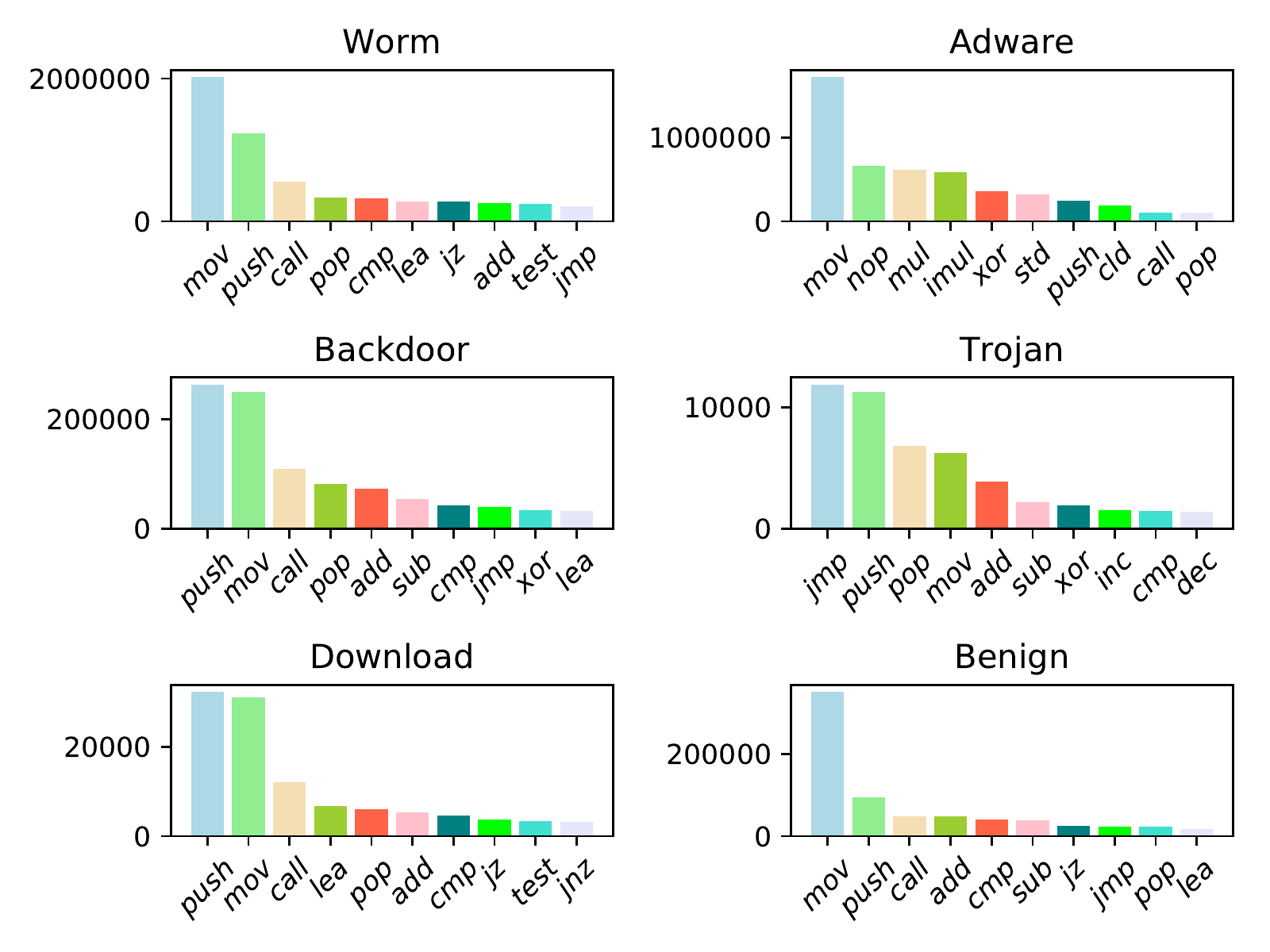}}
\caption{Most frequent 10 opcodes for each type of samples in the dataset.}
\label{fig}
\end{figure}

Based on statistical results, we observe that that most of the instructions are executed sequentially when the program are executed. That is, the jump instructions and function call instruction occupy only a small proportion. This process is very similar to natural language processing. Hence, as shown in Figure 6, we can make an analogy between a long article and an assembly instruction file. Then we also make an equivalent analogy between an assembly instruction function and a complete sentence in the article. Similarly, each instruction is analogous to the word in the sentence.
\begin{figure}[htbp]
\centerline{\includegraphics[width=8cm,height=6cm]{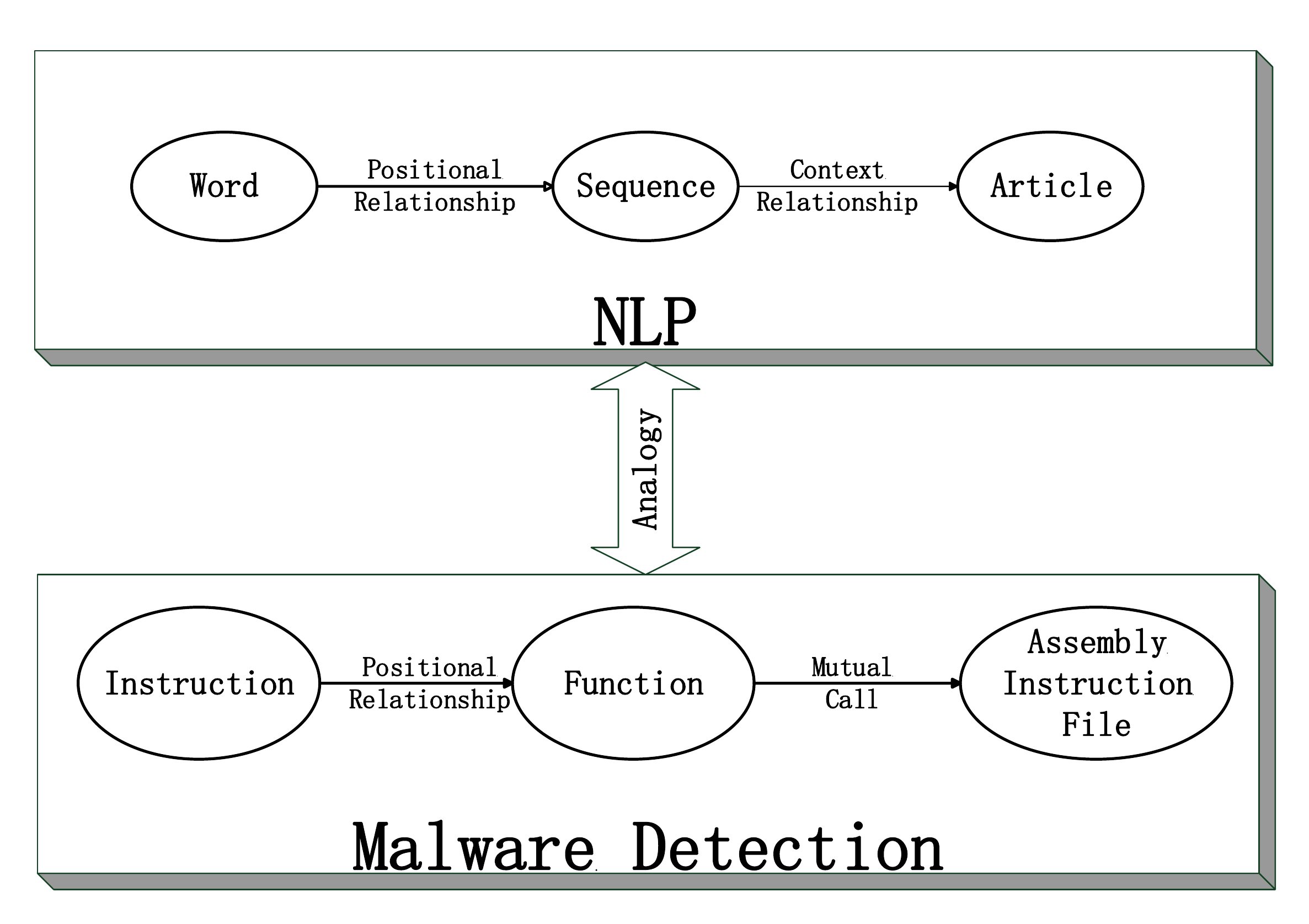}}
\caption{An equivalent analogy between NLP and malware detection.}
\label{fig}
\end{figure}

The positional relationship between words constitutes a sentence, and the context between sentences constitutes an article. Similarly, the positional relationship between instructions constitutes an assembly function, and the mutual call between functions forms an assembly instruction file. Although the instructions consist of opcodes and operands, in this paper, we only use opcodes that represent specific operational behaviors to replace instructions. In the end, the experimental results also show that this analogy intuition is indeed feasible.

\begin{figure*}[htbp]
\centerline{\includegraphics[width=18cm,height=9cm]{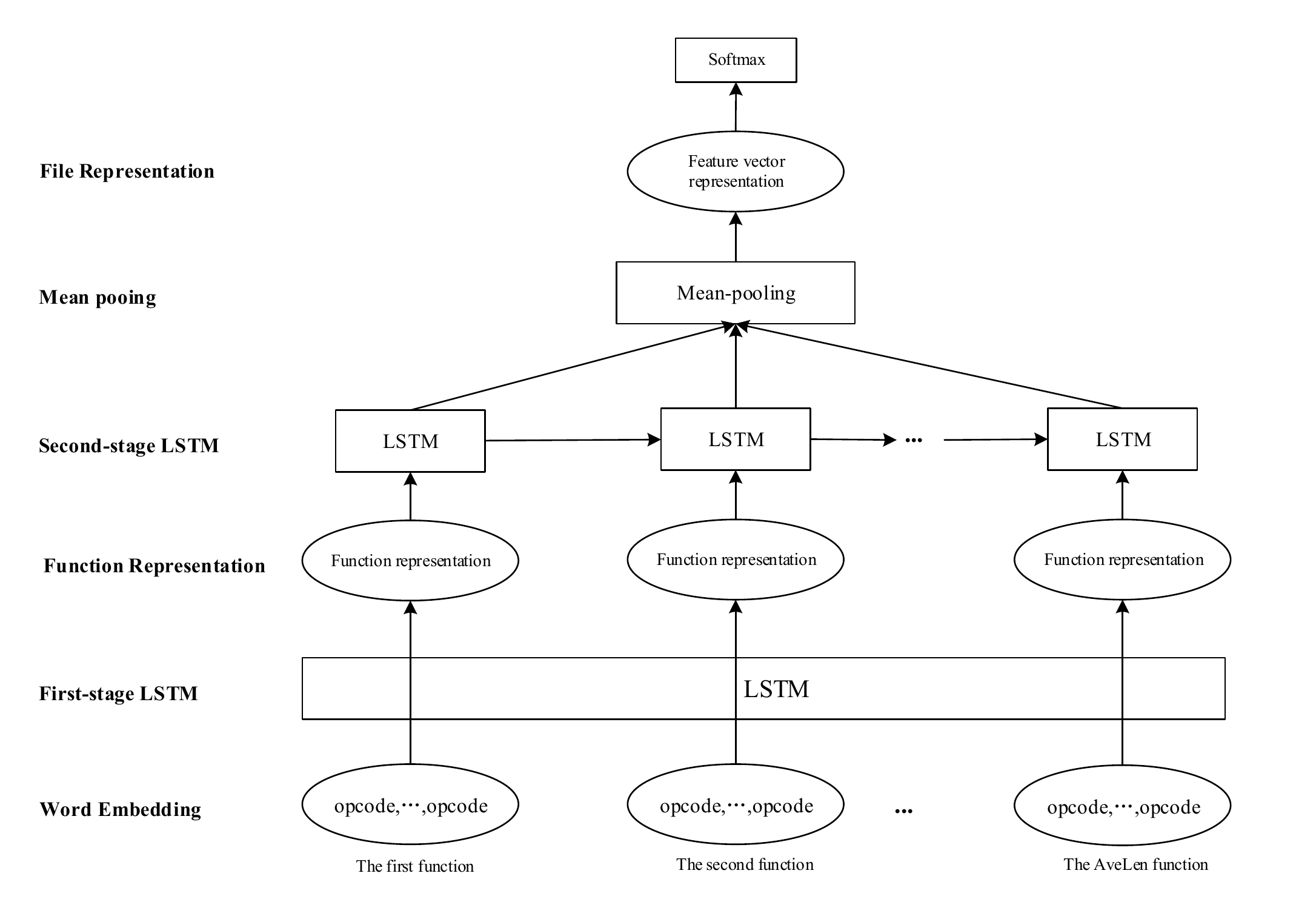}}
\caption{The model is consists of two LSTM hidden layer, a mean-pooling layer and a softmax layer.}
\label{fig}
\end{figure*}

Therefore, we propose a two-stage LSTM model for malware detection, which can handle opcode sequence of different length. Figure 7 shows the structure of the two-stage LSTM. We use two LSTM layers and one mean-pooling layer to obtain the feature representation of each opcode sequence file. In this paper, every opcode is represented as a 100-dimensional vector. The input of the first LSTM layer is all word embedding in the opcode sequence, and its output is corresponding function vector representation. Because LSTM current timestep input contains both the output of the previous timestep and the input of the current timestep, so we can use the output of the last timestep in each function to be as function vector representation. And all the function vectors are the input of the second LSTM layer. 

Differently, we added a mean-pooling layer after the second LSTM layer, which can enhance the invariance of feature representation. Assuming that the second LSTM layer will produce a representation sequence $h_1$, $h_2$, ..., $h_{AveLen}$, the mean-pooling layer will average over all timesteps resulting in representation $h_{average}$. That is, the hidden vectors obtained at every timestep in the second LSTM layer are averaged to acquire the feature representation of entire opcode sequence. Because of the use of this two-stage LSTM model, we can not only consider the positional relationship between opcodes, but also combine semantic information between functions to obtain a comprehensive feature representation of opcode sequence file. Finally, this representation is fed as the feature to a softmax layer to output the probability of classifying the executable file as every class. Softmax function is calculated as follows, where N is the number of executable file classes and $S_i$ is the probability of belonging to the i-th class.
\begin{equation}
S_i = \frac{exp(v_i)}{\sum_{j=1}^{N}exp(v_j)}
\end{equation}

\section{Experiments and Evaluations}
In this section, we present the experimental results to demonstrate the malware detection performance of our proposed method. We use python language to implement this malware detection framework based on Tensorflow \cite{r20} and Keras \cite{r21}, and all the experiments are conducted on a computer with 2.8GHz Intel Core i7 CPU, NVIDIA GeForce GTX 1050 GPU and 8GB memory.
\subsection{Data Collection}
Raw collections consist of 969 malware samples and 123 benign samples. Part of them are from the malware samples provided by Microsoft \cite{r22}. For each malware sample, two file formats are provided, the hexadecimal representation of the file's binary content and the corresponding assembly (\emph{.asm}) format files, but we only use the \emph{.asm} file generated by IDA Pro. The others are the executable file we collect, including benign applications such as browsers and system programs, as well as various malwares collected from some public malware websites such as MalwareDB \cite{r23} and Virusshare \cite{r24}. By reference to Microsoft’s classification format and Virustotal's analysis result \cite{r25}, these malwares are divided into the following categories: Worm, Adware, Backdoor, Trojan and Downloader. Then, we use the disassembly tool, IDA Pro, to get the corresponding \emph{.asm} file to form the entire dataset. The details on dataset can be seen Table 1. Finally, We randomly choose 70\% samples as training set and choose 30\% samples as testing set.

\begin{table}[!htbp]
\centering
\caption{The details on dataset}
\begin{tabular}{cccc}
\toprule
Collections & Numbers \\
\midrule
Worm & 155\\
Adware & 248\\
Backdoor & 442\\
Trojan & 48\\
Downloader & 76\\
Benign & 123\\
\bottomrule
\end{tabular}
\end{table}

\subsection{Evaluation Metric}
For the sake of convenience, we first introduce the confusion matrix for malware detection, as shown in Table 2.
\begin{table}[htbp]
\caption{Confusion matrix}
\begin{center}
\begin{tabular}{|c|c|c|c|}
\hline
\multicolumn{2}{|c|}{\multirow{2}*{}} & \multicolumn{2}{|c|}{Predicted class} \\
\cline{3-4}
\multicolumn{2}{|c|}{~} & Malware & Benign \\
\hline
\multirow{2}*{Actual} & Malware & TP & FN \\
\cline{2-4}
~ & Benign & FP & TN \\
\hline
\end{tabular}
\end{center}
\end{table}

In the evaluation section, we use the following metrics: true positive rate (TPR), false positive rate (FPR) and accuracy (ACC). TPR is equal to TP/(TP+FN), and FPR is equal to FP/(FP+TN), and ACC is equal to (TP+TN)/(TP+FN+FP+TN). An excellent model will have a higher TPR, a higher ACC, and a lower FPR. Hence, we also use receiver operating characteristic (ROC) curve in order to measure comprehensively both TPR and FPR. ROC curve typically features TPR on the Y axis, and FPR on the X axis. That is, the top left corner of the plot is the “ideal” point - a FPR of zero and a TPR of one so that a larger area under the curve (AUC) is usually better.

\subsection{Experimental Results}

We conduct lots of experiments to evaluate the dection performance of our proposed malware analysis method. The experiment is divided into two parts. On one hand, we collectively refer to Worm, Adware, Backdoor, Trojan, and Downloader as malware. In other words, it is a binary classification problem, either malware or benign file. On the other hand, we conducted more detailed malware classification experiments, which is a multi-classification problem. Anyway, we first use the training set to train this two-stage LSTM model (predictive model), and then use the trained model to predict the testing set. For the sake of generality, all of the models mentioned in this paper are used to predict the testing set after training 100 epochs using the traing set.
\subsubsection{Preferences}
First, we discuss the impact of different word window sizes and different word embedding techniques (Skip-gram and CBOW) on malware detection performance. Table 3 lists the experimental results. In general, the malware detection performance of using the CBOW model as word embedding techniqeu is better than the malware detection performance of using the Skip-gram model as word embedding technique. It can also be seen that setting word window size to 10 and choosing CBOW model as word embedding technique is the best option. Hence, we use the CBOW model as word embedding technique and set the word window size to 10 in the following experiments.
\begin{table*}[htbp]
\caption{Detection accuracy of different word window sizes and different word embedding techniques}
\begin{center}
\begin{tabular}{|c|c|c|c|c|}
\hline
\multirow{2}*{Word window size} & \multicolumn{2}{|c|}{Skip-gram (ACC)} & \multicolumn{2}{|c|}{CBOW (ACC)}\\
\cline{2-5}
~ & Binary-classification & Multi-classification & Binary-classification & Multi-classificatio\\
\hline
5 & 96.34\% & 92.07\% & 96.95\% & 89.94\% \\
\hline
7 & 96.95\% & 91.16\% & 96.95\% & 91.16\% \\
\hline
10 & 97.26\% & 92.99\% & 97.87\% & 94.51\% \\
\hline
15 & 97.18\% & 91.77\% & 97.56\% & 90.55\% \\
\hline
30 & 95.73\% & 91.46\% & 95.73\% & 92.38\% \\
\hline
\end{tabular}
\end{center}
\end{table*}

\subsubsection{The effect of the second LSTM layer}
Figure 8 and Figure 9 show the impact of the second LSTM layer on malware detection performance. It can be seen that the malware detection performance of two-stage LSTM model is always better than `not second LSTM layer' model for both binary classification problem and multi-classification problem.

\subsubsection{The comparison with other approachs}
Figure 10 and Figure 11 show the performance comparison of the two-stage LSTM model and other related malware detection approachs. Related approachs include the convolutional neural network (CNN), RNN and the multilayer perceptron (MLP). It can be seen that the performance of two-stage LSTM model is significantly better than RNN for both binary classification and multi-classification. It can also be seen that the two-stage LSTM model is better than MLP for both binary classification and multi-classification, and is slightly better than CNN for binary classification and multi-classification. Although the performance of CNN and two-stage LSTM model is comparable, the time cost of training CNN model is much greater than the time cost of training two-stage LSTM model because CNN model usually has a very large number of parameters.

Based on the experimental results, we can conclude that the method we proposed performs excellently on malware detection and malware classification.
\begin{figure}[htbp]
\centerline{\includegraphics[width=9cm,height=4cm]{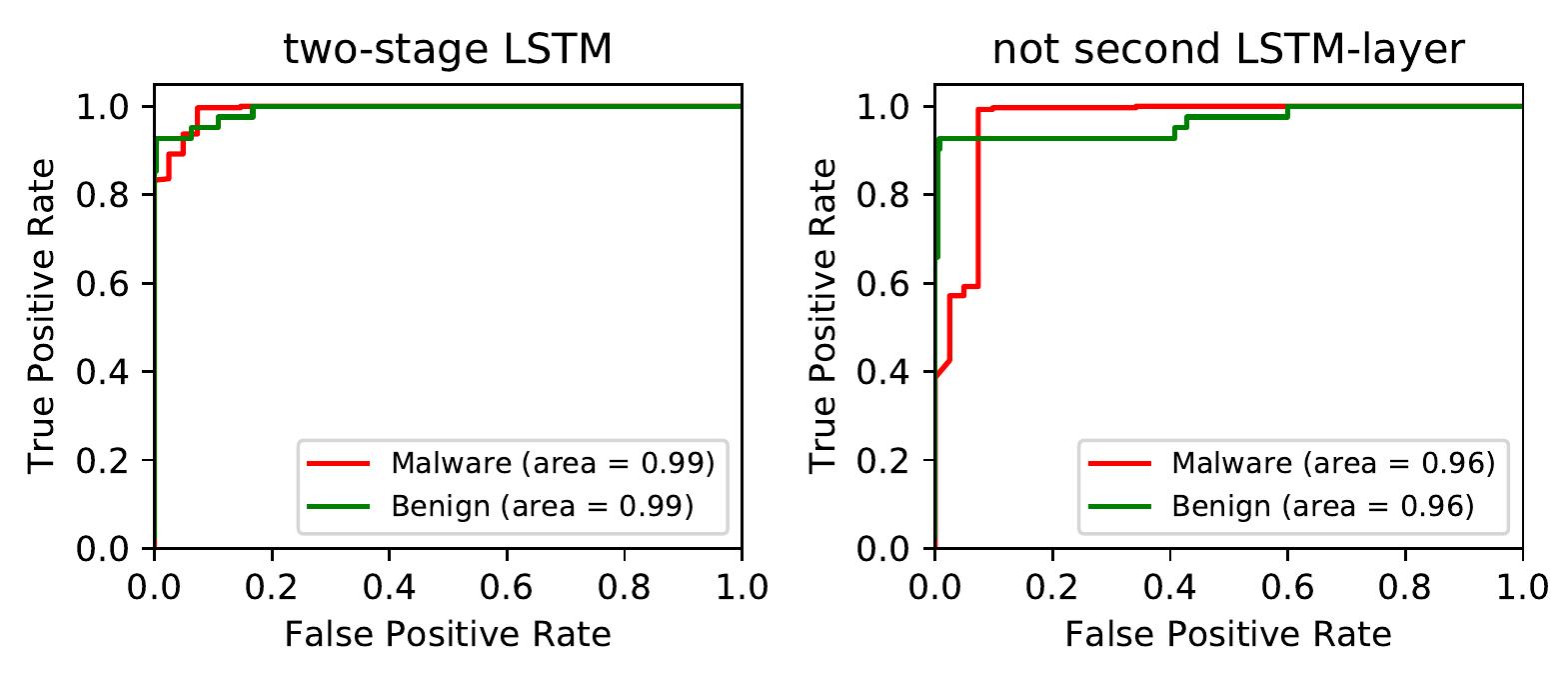}}
\caption{Binary classification: the performance comparison of two-stage LSTM model and `not second LSTM layer' model.}
\label{fig}
\end{figure}

\begin{figure}[htbp]
\centerline{\includegraphics[width=9cm,height=4cm]{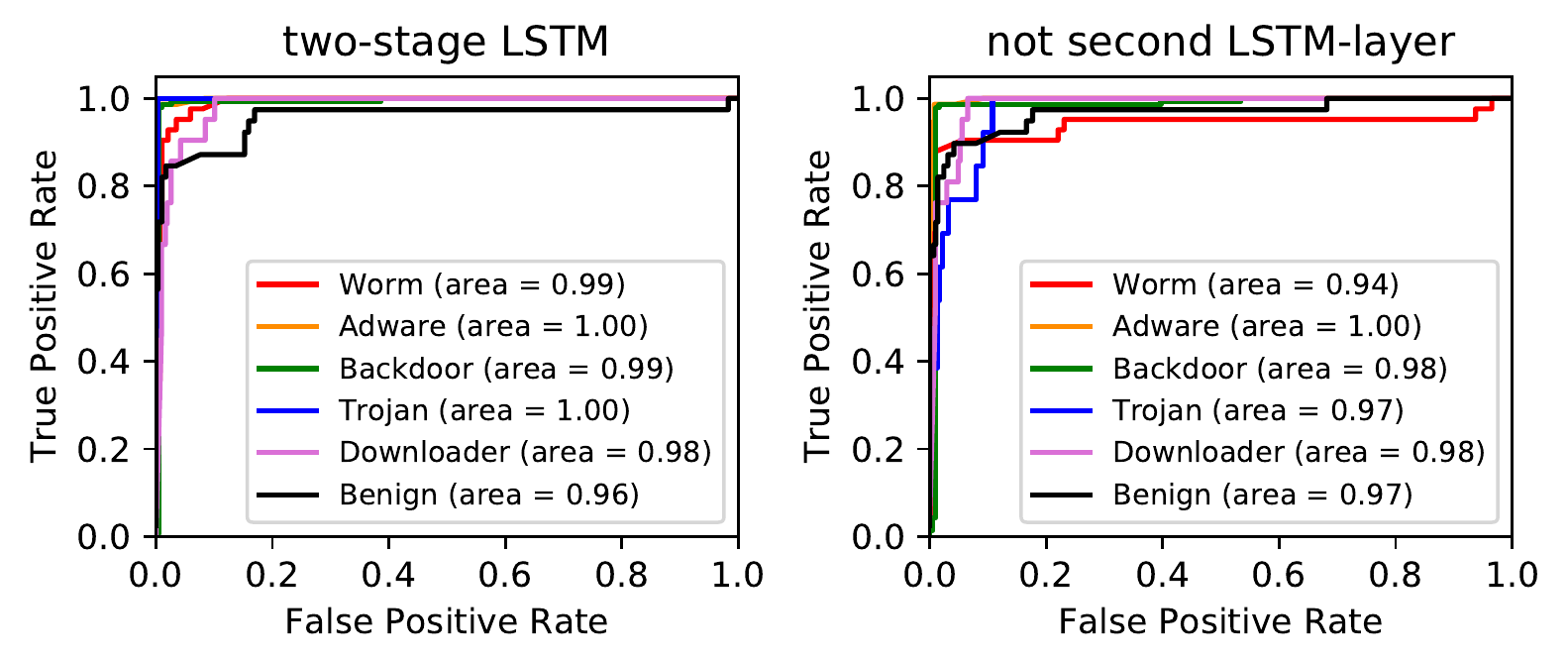}}
\caption{Multi-classification: the performance comparison of two-stage LSTM model and `not second LSTM layer' model.}
\label{fig}
\end{figure}

\begin{figure}[htbp]
\centerline{\includegraphics[width=9cm,height=6.5cm]{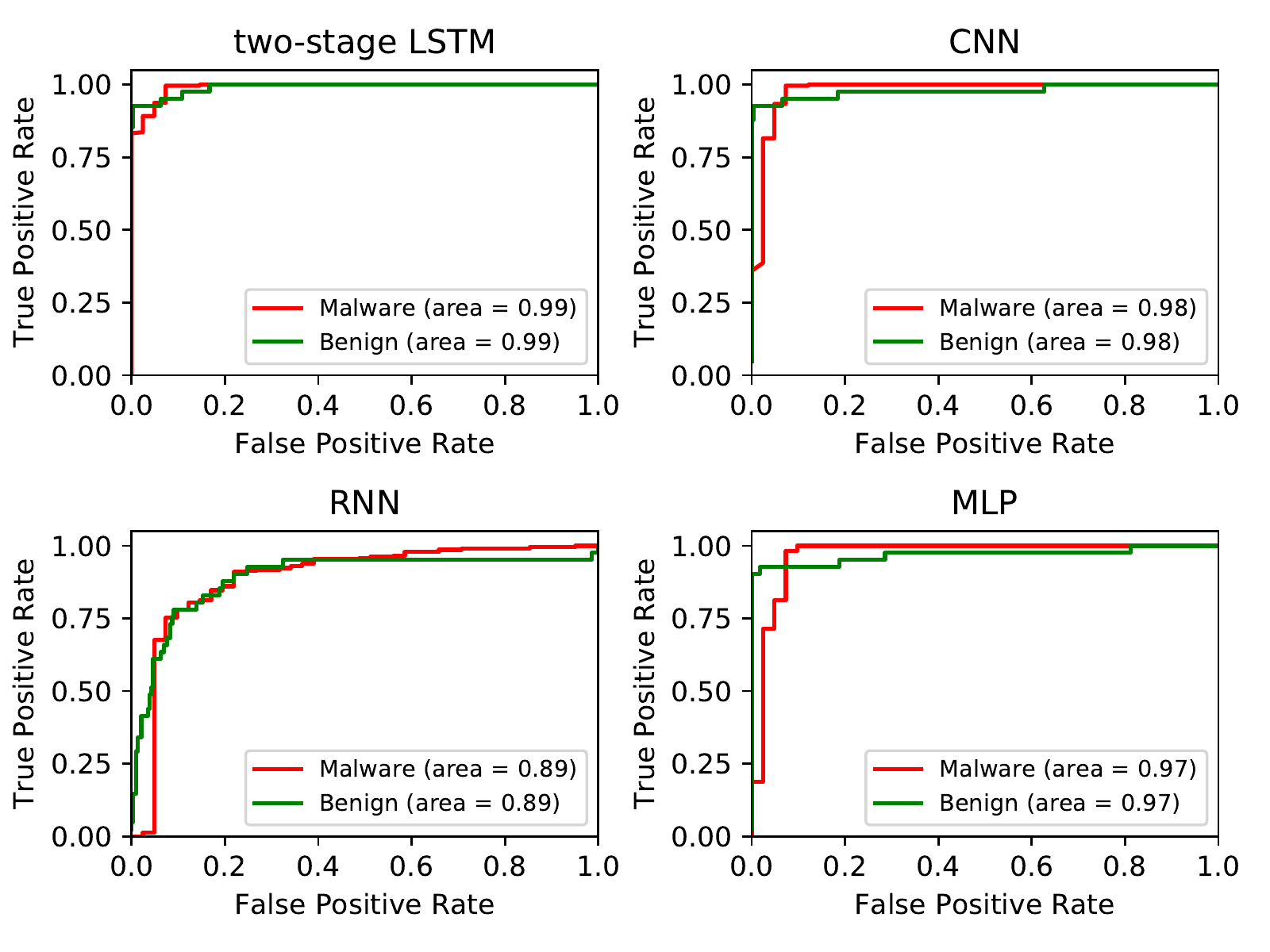}}
\caption{Binary-classification: the performance comparison of the two-stage LSTM model and other related malware detection approachs.}
\label{fig}
\end{figure}

\begin{figure}[htbp]
\centerline{\includegraphics[width=9cm,height=6.5cm]{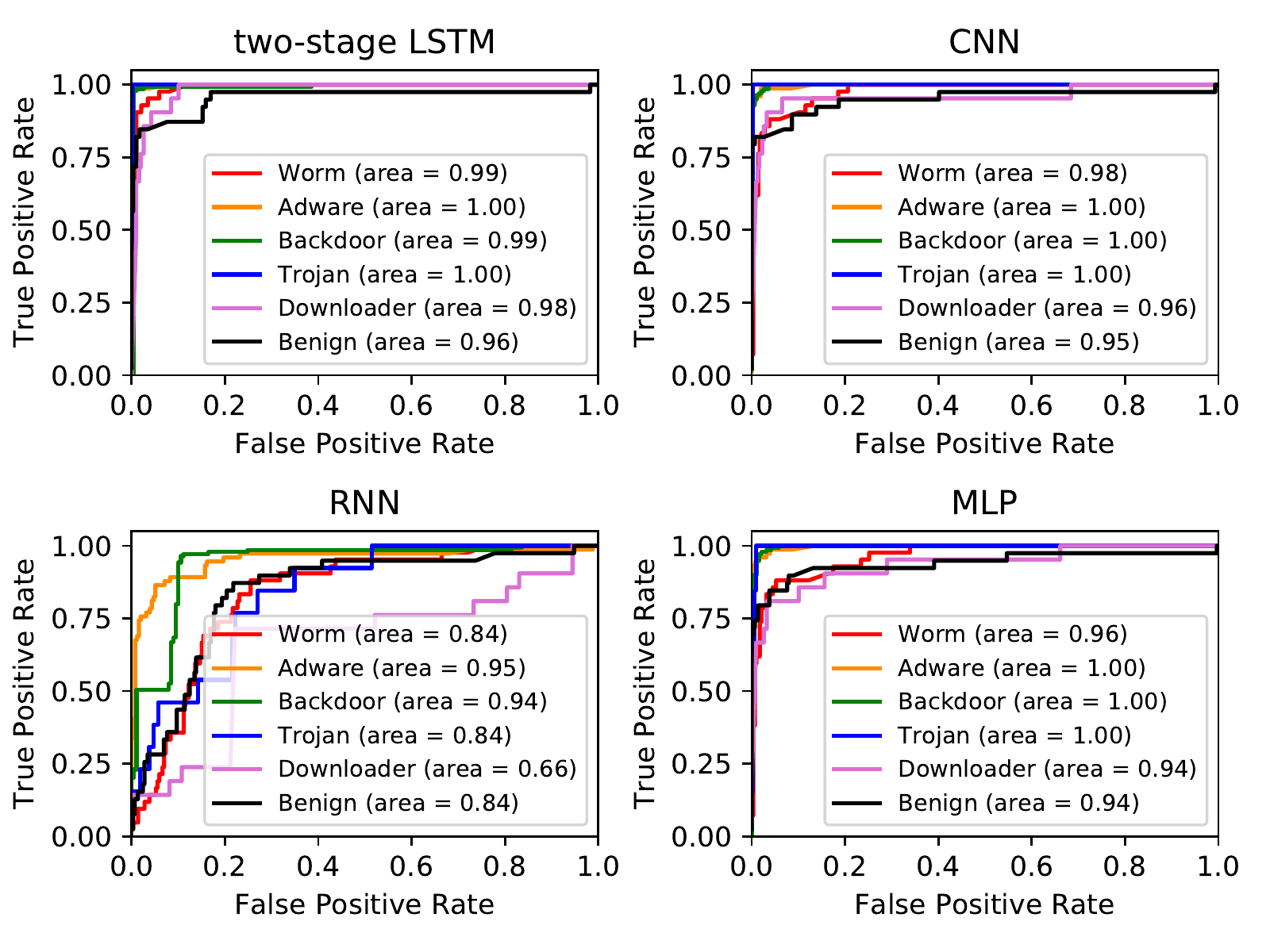}}
\caption{Multi-classification: the performance comparison of the two-stage LSTM model and other related malware detection approachs.}
\label{fig}
\end{figure}

\section{Conclusions}
In this paper, we propose a novel and efficient malware detection method. Similar to natural language processing, the word embedding technique and LSTM are used to automatically learn the correlation between opcodes and feature representation of opcode sequence, respectively. In addition, to increase invariance of the local feature representation, we also introduce a mean-pooling layer after second LSTM layer. To verify the availability of the proposed method, we perform a series of experiments on the dataset that includes 969 malwares and 123 benign files. Experimental results also demonstrate the effectiveness of our proposed method in malware detection and malware classification. 

On the other hand, the instructions consist of opcodes and operands. However, we only use the opcodes to conduct the experiment. Although the opcodes can reflect the specific operational behavior of the program, the operands may also reveal some different sensitive information between malware and benign files. Therefore, the use of operands for malware detection will be our future work.

\bibliographystyle{IEEEtran}
\bibliography{ref}

\end{document}